\input harvmac\skip0=\baselineskip
\input epsf

\newcount\figno
\figno=0
\def\fig#1#2#3{
\par\begingroup\parindent=0pt\leftskip=1cm\rightskip=1cm\parindent=0pt
\baselineskip=11pt \global\advance\figno by 1 \midinsert
\epsfxsize=#3 \centerline{\epsfbox{#2}} \vskip 12pt {\bf Fig.\
\the\figno: } #1\par
\endinsert\endgroup\par
}
\def\figlabel#1{\xdef#1{\the\figno}}
\def\encadremath#1{\vbox{\hrule\hbox{\vrule\kern8pt\vbox{\kern8pt
\hbox{$\displaystyle #1$}\kern8pt} \kern8pt\vrule}\hrule}}



\lref\osv{
  H.~Ooguri, A.~Strominger and C.~Vafa,
  ``Black hole attractors and the topological string,''
  Phys.\ Rev.\ D {\bf 70}, 106007 (2004)
  [arXiv:hep-th/0405146].
}

\lref\WittenKT{
  E.~Witten,
  ``Three-Dimensional Gravity Revisited,''
  arXiv:0706.3359 [hep-th].
}

\lref\Tuite{
  M.~P.~Tuite,
  ``Genus Two Meromorphic Conformal Field Theory,''
  arXiv:math.qa/9910136.
}

\lref\Igusa{
  J.-I.~Igusa,
  ``On Siegel Modular Forms of Genus Two,"
  Am.J.Math. {\bf 84} (1962) 175-200;
  ``Modular Forms and Projective Invariants,"
  Am.J.Math. {\bf 89} (1967) 817-855.
}

\lref\FLM{ I.~Frenkel, J.~Lepowsky and A.~Meurman, ``A Natural
Representation of the Fischer-Griess Monster with the Modular
Function J as Character,'' Proc.Natl.Acad.Sci.USA {\bf 81} (1984)
3256-3260;
  I.~Frenkel, J.~Lepowsky and A.~Meurman,
  ``Vertex Operator Algebras and the Monster,''
{\it  Boston, USA: Academic (1988) 508 P. (Pure and Applied
Mathematics, 134)} }

\lref\ZamolodchikovAE{
  A.~B.~Zamolodchikov, ``Conformal Scalar Field on the
  Hyperelliptic Curve and Critical Ashkin-Teller Multipoint
  Correlation Functions,'' Nucl.\ Phys.\  B {\bf 285}, 481 (1987).
}

\lref\KnizhnikXP{
  V.~G.~Knizhnik,
  ``Analytic Fields on Riemann Surfaces. 2,''
  Commun.\ Math.\ Phys.\  {\bf 112}, 567 (1987).
}

\lref\DixonQV{
  L.~J.~Dixon, D.~Friedan, E.~J.~Martinec and S.~H.~Shenker,
  ``The Conformal Field Theory Of Orbifolds,''
  Nucl.\ Phys.\  B {\bf 282}, 13 (1987).
}

\lref\HamidiVH{
  S.~Hamidi and C.~Vafa,
  ``Interactions on Orbifolds,''
  Nucl.\ Phys.\  B {\bf 279}, 465 (1987).
}

\lref\Hohn{ G.~Hoehn, ``Selbstduale Vertexoperatorsuperalgebren und
das Babymonster,'' arXiv:0706.0236. }

\lref\Tuiteb{ G.~Mason and M.~P.~Tuite, ``On Genus Two Riemann
Surfaces Formed from Sewn Tori,'' arXiv:math.qa/0603088. }

\Title{\vbox{\baselineskip12pt\hbox{} }} {\vbox{\centerline{Genus Two Partition Functions of}
\vskip12pt\centerline{Extremal Conformal
Field Theories}}} \centerline{ Davide Gaiotto and Xi Yin }
\smallskip
\centerline{Jefferson Physical Laboratory, Harvard University,
Cambridge, MA 02138} \vskip .6in \centerline{\bf Abstract} {
Recently Witten conjectured the existence of a family of ``extremal"
conformal field theories (ECFTs) of central charge $c=24k$, which
are supposed to be dual to three-dimensional pure quantum gravity in
$AdS_3$. Assuming their existence, we determine explicitly the genus
two partition functions of $k=2$ and $k=3$ ECFTs, using modular
invariance and the behavior of the partition function in
degenerating limits of the Riemann surface. The result passes highly
nontrivial tests and in particular provides a piece of evidence for
the existence of the $k=3$ ECFT. We also argue that the genus two
partition function of ECFTs with $k\leq 10$ are uniquely fixed (if they
exist). } \vskip .3in

\Date{July 2007}

\listtoc \writetoc \noblackbox

\newsec{Introduction}

Recently Witten \WittenKT\ argued that pure three-dimensional
quantum gravity with a negative cosmological constant in $AdS_3$
should be dual to a CFT on the boundary of central charge
$(c_L,c_R)=(24k,24k)$, where $k$ is a positive integer. This CFT
factorizes into a holomorphic CFT and an anti-holomorphic CFT, whose
lowest dimensional primary field has dimension $k+1$. Such CFTs are
called extremal (ECFT) \Hohn. A $k=1$ ECFT was constructed by
Frenkel, Lepowsky and Meurman \FLM\ as a ${\bf Z}_2$ orbifold of
free bosons on the Leech lattice, giving rise to the monster module.
It is not yet known whether the $k>1$ ECFTs exist, and it is clearly
of interest to either construct them or to disprove their existence.

It was shown in \WittenKT\ that the partition function for a $k=2$
ECFT, if exists, can be constructed on any genus $g$ hyperelliptic
Riemann surface, using the $(2g+2)$-point function of twist
operators in the 2-fold symmetric product of the $k=2$ ECFT. The
partition function constructed in this way is consistent in the
sense that, in the limit where the Riemann surface degenerates, the
partition function reduces to lower genus correlation functions in
suitable ways.

For example, the genus one partition function is related to the
four-point function of the twist field ${\cal E}$. The latter can be
determined from the ${\cal E}(z) {\cal E}(0)$ OPE, which essentially
encodes the operator spectrum of the CFT. The genus two partition
function, on the other hand, is related to six-point function of
twist fields, and encodes information about the three point function
of primaries. It is an $Sp(4,{\bf Z})$ modular form of weight $2k$,
and is in fact $\chi_{10}^{-k}$ times an entire holomorphic Siegel
modular form of weight $12k$ \WittenKT. Here $\chi_{10}$ stands for
the weight 10 Igusa cusp form. The $k=1$ genus two partition
function has been computed in \Tuite. For $k=1,2,3$, there are a
basis of $3$, $8$ and $17$ linearly independent entire $Sp(4,{\bf
Z})$ modular forms of weight $12$, $24$ and $36$, respectively. One
can determine the coefficients of these basis modular forms by
considering the limits where the genus two Riemann surface
degenerates. One limit (``pairwise degeneration") is when a handle
of the Riemann surface is pinched, corresponding to a pair of the
twist fields collide. Another limit (``separating degeneration") is
when the Riemann surface degenerates into two genus one surfaces
touching at a point (or conformally equivalently, connected by a
thin tube).

In practice one can simplify things by considering a limit where all three pairs of twist fields degenerate,
so that the six-point function can be replaced by the three point function of operators
appearing in the singular terms of the ${\cal E}(z){\cal E}(0)$ OPE. The latter is of the form
\eqn\eeopee{ {\cal E}(z) {\cal E}(0) = {1\over z^{3k}}(1+{\rm Virasoro~descendants})
+ {1\over z^{k-2}}\sum_i {\cal O}_{k+1,i}^+ {\cal O}_{k+1,i}^- +\cdots }
where ${\cal O}_{k+1,i}^\pm$ are primaries of dimension $k+1$ in the two copies of the ECFT.
Without using any information of ${\cal O}_{k+1,i}$, one can determine certain singular parts of
the six-point function of ${\cal E}$. This turns out to be sufficient to fix (in fact, ``over"-determining) the $k=1$ and $k=2$
genus two partition functions completely, while for $k=3$ one can fix all but 3 linear combinations of the
17 coefficients of the Siegel modular forms.

On the other hand, at the separating degeneration, the leading divergence of the
genus two partition function factorizes as the product of the partition functions
of the two genus one Riemann surfaces. For $k=1,2$, this is indeed the case, as expected from \WittenKT. It also
provides a highly nontrivial check for our expression for the $k=2$ genus two partition function.
For $k=3$, the factorization at the separating degeneration fixes the remaining 3 coefficients of
the modular forms, and hence the genus two partition function. Once again, it in fact ``over"-determines
the genus two partition function, hence the consistent factorization of our expression
provides nontrivial evidences for the existence of the $k=3$ ECFT.

A slightly different approach, suggested in \WittenKT\ as well, is
to start by constructing the genus two partition function by sewing
two genus one Riemann surfaces. Combining the knowledge of certain
torus one-point functions and the $Sp(4,{\bf Z})$ modular
invariance, we will show that it is in fact possible (at least in
principle) to fix the genus two partition functions completely (and
uniquely), for ECFTs with $k\leq 10$, assuming their existence. The
explicit solutions, as well as consistency checks at all
degenerations of the genus two Riemann surface, will be left to
future work.

Section 2 describes some useful formulae for the OPE of twist fields
and Siegel modular forms. In section 3, we shall examine the
partition function of ECFTs with $k=1,2,3$. In section 4, we discuss
the factorization at the separating degeneration to higher orders,
and a general sewing construction of the genus two partition
function.

\newsec{Generalities}

The 1-loop partition function of an extremal CFT $M$ can be related
to the 4-point function of twistor operators in the symmetric
product CFT ${\rm Sym}^2(M)$
\refs{\ZamolodchikovAE,\KnizhnikXP,\DixonQV,\HamidiVH,\WittenKT},
\eqn\zta{ Z = 2^{8k} \left(\prod_{1\leq i<j\leq 4} e_{ij}\right)^k
\langle {\cal E}(e_1) {\cal E}(e_2) {\cal E}(e_3) {\cal E}(e_4)
\rangle. } where ${\cal E}$ is normalized such that $\langle{\cal
E}(x){\cal E}(0)\rangle=x^{-3k}$. The $e_i$'s are related to the
modulus $\tau$ of the torus as follows. If we set $e_4=\infty$,
$e_1+e_2+e_3=0$ by a conformal transformation, then the Jacobi theta
functions of $\tau$ are related by \eqn\thetae{
e_{12}=\theta_3(\tau)^4,~~~~ e_{32} = \theta_2(\tau)^4, ~~~~
e_{13}=\theta_4(\tau)^4. } For general $e_i$'s, we can write the
$j$-function of $\tau$ as \eqn\jgen{ j(\tau(e_1,e_2,e_3,e_4)) = 2^5
{(\theta_2^8+\theta_3^8+\theta_4^8)^3\over \theta_2^8 \theta_3^8
\theta_4^8}= 2^5
{(e_{12}^2e_{34}^2+e_{13}^2e_{24}^2+e_{14}^2e_{23}^2)^3\over
(\prod_{i<j} e_{ij})^2} } where $e_{ij}\equiv e_i-e_j$.

The OPE of the twist fields of the form \eqn\eeope{ {\cal E}(x)
{\cal E}(0)\sim {1\over x^{3k}}(1+{\rm descendants}) + {1\over
x^{k-2}} {\cal O}_{2k+2}(0)+\cdots } where ${\cal O}_{2k+2}$ is a
primary field of dimension $2k+2$ in the untwisted sector of ${\rm
Sym}^2(M)$. By examining the three-point function $\langle {\cal
E}{\cal E}{\cal O}\rangle$ one can see that ${\cal O}_{2k+2}$ is in
fact  proportional to $\sum_i {\cal O}_{k+1,i}^+ {\cal
O}_{k+1,i}^-$, where ${\cal O}_{k+1,i}^\pm$ are the complete set of
dimension $k+1$ primaries in the two copies of the ECFTs.

To determine the Virasoro descendants appearing in \eeope, we shall
closely follow the discussion of \WittenKT, but will work to higher
orders. Inserting a pair of twist fields ${\cal E}(e)$, ${\cal
E}(-e)$ in a correlation function amounts to compute the correlation
function on the covering Riemann surface $y^2=(x+e)(x-e)$. The
Virasoro descendants appearing in the RHS of \eeope\ can be
determined by requiring that the corresponding state is annihilated
by the difference of the Virasoro generators on the two branches of
the covering Riemann surface. Let $u=x+y, v=x-y$. The equation
defining the double cover of the $x$-plane branched at $\pm e$ is
then $uv=e^2$. The holomorphic vector fields \eqn\vnn{ V_n = 2^{-n}
u^{n+1}\partial_u = -2^{-n} e^{2n} v^{1-n}\partial_v } define the
Virasoro generators \eqn\virag{ Q_n^+ = \oint_{S_+} V_n T =
\oint_{S_+} 2^{-n} u^{n+1}{dx\over du} dx T_{xx} } on the upper
sheet, and \eqn\virag{ Q_n^- = \oint_{S_-} V_n T = \oint_{S_-}
2^{-n} e^{2n} u^{1-n}{dx\over du} dx T_{xx} } on the lower sheet, up
to a constant term due to the anomaly in transforming $T$ from $u$
to $x$ coordinate. The operators $\hat Q_n=Q_n^+-Q_n^-$ should
annihilate the state appearing in the ${\cal E}(e){\cal E}(-e)$ OPE.
The constant terms in the $\hat Q_n$'s can be determined by
requiring $[\hat Q_n,\hat Q_m]=(n-m) \hat Q_{n+m}$. For our purpose,
we will need the expressions for $\hat Q_{0,1,2,3,4}$ up to terms of
order ${\cal O}(e^8)$. They are given explicitly by
\eqn\qss{\eqalign{ & \hat Q_0 = \left( L_0^+ - {e^2\over 2}L_{-2}^+
-{e^4\over 8} L_{-4}^+ -{e^6\over 16}L_{-6}^+ -{5e^8\over
128}L_{-8}^+ \right) \cr &~~~~~~~~~-\left( L_0^- - {e^2\over
2}L_{-2}^- -{e^4\over 8} L_{-4}^- -{e^6\over 16}L_{-6}^- -{5e^8\over
128}L_{-8}^- \right)+\cdots \cr & \hat Q_1 = \left( L_1^+ -
{3e^2\over 4} L_{-1}^+ - {e^4\over 16} L_{-3}^+ -{e^6\over 32}
L_{-5}^+ - {5e^8\over 256} L_{-7}^+ \right) \cr &~~~~~~~~~ - \left(
{e^2\over 4}L_{-1}^- - {e^4\over 16} L_{-3}^- - {e^6\over 32}
L_{-5}^- - {5e^8\over 256}L_{-7}^- \right)+\cdots \cr & Q_2 = \left(
L_2^+ - e^2 L_0^+ -3ke^2 + {e^4\over 16} L_{-2}^+ - {e^8\over 256}
L_{-6}^+ \right) - \left( {e^4\over 16} L_{-2}^- - {e^8\over 256}
L_{-6}^- \right) +\cdots \cr & \hat Q_3 = \left( L_3^+ - {5e^2\over
4} L_1^+ + {e^4\over 4}L_{-1}^+ + {e^6\over 64}L_{-3}^+ + {e^8\over
256} L_{-5}^+ \right) - \left( {e^6\over 64}L_{-3}^- + {e^8\over
256} L_{-5}^- \right)+\cdots \cr & \hat Q_4 = \left( L_4^+
-{3e^2\over 2}L_2^+ + {e^4\over 2} L_0^+ - {3ke^4\over 2} +
{e^8\over 256} L_{-4}^+
 \right) - {e^8\over 256} L_{-4}^- +\cdots  }} where $L^{\pm}$ are the Virasoro generators in the two
 copies of the CFT, i.e. on the two sheets.
The state $|\Psi\rangle$ of the form $(1+{\rm
descendants})|0\rangle$ and annihilated by all the $\hat Q_m$'s is
\eqn\psit{\eqalign{  |\Psi\rangle &= \left\{ 1 + {e^2\over 4} L_{-2}
+ {e^4\over 32} \left[L_{-4} + L_{-2}^2 + {1\over 6k} L_{-2}^+
L_{-2}^- \right]\right. \cr &\left. + {e^6\over
128}\left[L_{-2}L_{-4} +{1\over 6k} L_{-2}^+ L_{-2}^- L_{-2} +
{1\over 3} L_{-2}^3 + {1\over 24k} L_{-3}^+ L_{-3}^-\right]
\right.\cr &+ {e^8\over 512} \left[ {7\over 6} L_{-8} + {1\over 2}
(L_{-2}L_{-6}+L_{-6}L_{-2}) +{1\over 4} L_{-4}^2 + {1\over 6}
(L_{-2}^2 L_{-4}+L_{-2}L_{-4}L_{-2}+L_{-4}L_{-2}^2) \right.\cr & +
{1\over 12}L_{-2}^4 + {1\over 12k}L_{-4}L_{-2}^+L_{-2}^- +{1\over 12
k}L_{-2}^2L_{-2}^+L_{-2}^- + {1\over 24k}L_{-2} L_{-3}^+L_{-3}^- \cr
&\left.\left. + {1\over k(60k+11)} \left( (k+{1\over
3})L_{-4}^+L_{-4}^- + {5\over 12} (L_{-2}^+)^2(L_{-2}^-)^2 - {1\over
4}(L_{-4}^+(L_{-2}^-)^2 +L_{-4}^-(L_{-2}^+)^2) \right)
\right]\right. \cr &\left. + {\cal O}(e^{10}) \right\}|0\rangle }}
The corresponding operator is ($e=x/2$) \eqn\opet{\eqalign{ \Psi_x
&= 1+ {x^2\over 16} T + {x^4\over 2^{10}} \partial^2 T + {x^4\over
2^9} T*T + {x^4\over 3 \cdot 2^{10}k}T^+ T^- +{x^6\over
2^{14}}T*\partial^2 T  \cr & + {x^6\over 3\cdot
2^{14}k}(T^+*T^+T^-+T^-*T^-T^+) + {x^6\over 3\cdot 2^{13}} T*(T*T) +
{x^6\over 3\cdot 2^{16} k}\partial T^+ \partial T^- \cr & +{x^8\over
2^{17}} \left\{ {7\over 6\cdot 6!} \partial^6 T + {1\over 48}
(T*\partial^4 T+\partial^4T*T) +{1\over 16} \partial^2T*\partial^2T
+ {1\over 12}T*(T*(T*T)) \right.\cr &+ {1\over 12}
(T*(T*\partial^2T)+T*(\partial^2T*T)+\partial^2 T*(T*T))  + {1\over
24k}\partial^2T*(T^+T^-)\cr & +{1\over 12 k}T*(T*(T^+T^-)) + {1\over
24k}T*(\partial T^+\partial T^-) + {1\over 4k(60k+11)} \left[
({k}+{1\over 3})\partial^2T^+\partial^2T^- \right.\cr & \left.
\left.+ {5\over 3} (T^+*T^+)(T^-*T^-) - {1\over 2}(\partial^2 T^+
T^-*T^-+\partial^2T^- T^+*T^+) \right]  \right\}+ {\cal O}(x^{10})
}} where the notation $A*B$ stands for ${\rm Res}_{z\to 0}\left[A(z)
B(0)/z\right]$. Now we can express the OPE of twist fields as
\eqn\fpe{ {\cal E}(x/2) {\cal E}(-x/2) \sim {1\over x^{3k}}\Psi_x(0)
+ {const\over x^{k-2}} \sum_i {\cal O}_i^+ {\cal O}_i^-(0)+\cdots }

Let us consider the six-point function $\langle {\cal E}(e_1)\cdots
{\cal E}(e_6) \rangle$. It is related to the genus two partition
function by \eqn\genut{ Z_{k,g=2}(\Omega) = A_k \left[\prod_{1\leq
i<j\leq 6}(e_i-e_j)^k\right] \langle {\cal E}(e_1)\cdots {\cal
E}(e_6) \rangle } where the genus two Riemann surface is represented
as the hyperelliptic curve \eqn\hyperss{ y^2 = \prod_{i=1}^6
(x-e_i), } and $A_k$ is a constant. $Z_{k,g=2}$ is an ${\rm
Sp}(4,{\bf Z})$ modular form of weight $2k$. The six-point function
has singularities that goes like $(e_i-e_j)^{-3k}$. Multiplying it
by $\prod_{i<j}(e_i-e_j)^{3k}=\chi_{10}^{3k/2}$, one obtains an
entire holomorphic $Sp(4,{\bf Z})$ Siegel modular form of weight
$12k$. The ring of such modular forms is generated by the Eisenstein
series $\psi_4, \psi_6$ (with slightly different normalization, as
defined below) and the cusp forms $\chi_{10}, \chi_{12}$. Following
\Igusa, one defines the projective invariants \eqn\basiss{ \eqalign{
& A = \sum_{15~{\rm perms}} e_{12}^2 e_{34}^2 e_{56}^2, \cr & B =
\sum_{10 ~{\rm perms}} e_{12}^2 e_{23}^2 e_{31}^2 e_{45}^2 e_{56}^2
e_{64}^2, \cr & C = \sum_{60~{\rm perms}} e_{12}^2 e_{23}^2 e_{31}^2
e_{45}^2 e_{56}^2 e_{64}^2 e_{14}^2 e_{25}^2 e_{36}^2, \cr & D =
\prod_{1\leq i<j\leq 6} e_{ij}^2, } } There is a ring homomorphism
mapping Siegel modular forms to projective invariants. We can write
the projective invariants corresponding to generating modular forms
as (by an abuse of notation, we shall not distinguish the two)
\eqn\siegelgen{ \eqalign{ & \psi_4 = B, \cr & \psi_6={1\over
2}(AB-3C), \cr & \chi_{10}=D, \cr & \chi_{12}=AD. } } An important
property of the Siegel modular form is its factorization at the
separating degeneration of the genus two Riemann surface, where the
off-diagonal component $\tau_{12}$ of the period matrix goes to
zero. We shall use the parameter $\epsilon$ defined in
\refs{\Tuite,\Tuiteb}, related by $2\pi i\tau_{12} =-\epsilon+{\cal
O}(\epsilon^3)$. In the $\epsilon\to 0$ limit, \eqn\degensie{
\eqalign{ & \psi_4 = {1\over 4}E_4(\tau_1) E_4(\tau_2) + {\cal
O}(\epsilon^2),\cr & \psi_6= {1\over 16}E_6(\tau_1)
E_6(\tau_2)+{\cal O}(\epsilon^2), \cr & \chi_{10}= const\cdot
\epsilon^2 \Delta(\tau_1) \Delta(\tau_2) + {\cal O}(\epsilon^4),\cr
&\chi_{12} = 96 \Delta(\tau_1)\Delta(\tau_2)+{\cal O}(\epsilon^2). }
}

\newsec{Explicit results}

\subsec{The $k=1$ extremal CFT}

As a warm up exercise we shall revisit the genus one and genus two
partition functions of the $k=1$ extremal CFT. The
genus one partition function is \eqn\partko{ Z_1(q) = J(q), } where
$J(q)=j(q)-744$.  Identifying the four point function with \partko, we can expand the part of
$\langle {\cal E}(x/2) {\cal E}(-x/2) {\cal E}(y/2+z) {\cal
E}(-y/2+z) \rangle$ that is singular in $x,y$ in powers of $z$,
\eqn\foup{\eqalign{ & \langle {\cal E}(x/2) {\cal E}(-x/2) {\cal
E}(y/2+z) {\cal E}(-y/2+z) \rangle \cr &~~~~~~~~~~= x^{-3}y^{-3}+
{3\over 32}x^{-1}y^{-1}z^{-4}+{3\over
64}(xy^{-1}+x^{-1}y)z^{-6}+\cdots }}
This expression can indeed be reproduced from \fpe\ by explicitly evaluating
the two point function $\langle\Psi_x(0) \Psi_y(z)\rangle$.

The $k=1$ genus two partition function is a linear combination of
\eqn\thee{ {\psi_4^3\over \chi_{10}},~~~~{\psi_6^2\over
\chi_{10}},~~~~{\chi_{12}\over \chi_{10}}. } In the limit
$e_{12},e_{34},e_{56}\to 0$, the singular terms in the six-point
function of ${\cal E}(e_i)$ can be determined using the ${\cal
E}{\cal E}$ OPE \fpe, together with the three point function of
terms up to order $x^2$ in $\Psi_x$.

 By matching with these, one can fix the
unique choice of the modular form (up to overall normalization),
\eqn\thea{ Z_{k=1,g=2}(\Omega) = {A_1\over\chi_{10}}\left( {41\over
4608} \psi_4^3+{31\over 1152} \psi_6^2 - {3813\over 2048} \chi_{12}
\right)  } This is indeed the same expression as in \Tuite\ (note
the different convention for the generating forms: in \Tuite\
$F_{12}$ is not a cusp form; it is a more general linear combination of
$\chi_{12}$, $\psi_4^3$ and $\psi_6^2$). In the limit $\epsilon\to
0$, one can check that \thea\ indeed factorizes as
\eqn\facthea{Z_{k=1,g=2}(\Omega) \to {const\over \epsilon^2}
J(\tau_1) J(\tau_2).  } Note that $\Delta=(E_4^3-E_6^2)/1728$, and
$J=(41 E_4^3+31 E_6^2)/(72 \Delta)$.

We can extract information about the three-point functions of primaries from the genus two partition function \thea.
For example, by expanding the six-point function \eqn\sixpo{\langle {\cal E}({x\over 2}){\cal E}(-{x\over 2}){\cal E}({y\over 2}+u){\cal E}(-{y\over 2}+u)
{\cal E}({z\over 2}+v){\cal E}(-{z\over 2}+v) \rangle} corresponding to \thea, up to order
${\cal O}(x^{-3} y z)$ and ${\cal O}(xyz)$ respectively, and subtracting the contribution
from the Virasoro descendants in $\Psi_x$,\foot{The three-point functions of various
Virasoro descendants are rather messy, and are computed using a Mathematica program. The program is
available upon request. }
one obtains $\sum_{i,j} \langle {\cal O}_i {\cal O}_j \rangle^2$
and $\sum_{i,j,k} \langle {\cal O}_i {\cal O}_j {\cal O}_k\rangle^2$, where ${\cal O}_i$ are the 196883 dimension
2 primaries. Normalizing the ${\cal O}_i$'s such that $\langle {\cal O}_i(z) {\cal O}_j(0)\rangle = \delta_{ij}z^{-4}$, we find
\eqn\thept{ {1\over 196883}\sum_{i,j,k=1}^{196883} \langle{\cal O}_i(z_1) {\cal O}_j(z_2) {\cal O}_k(z_3)\rangle^2
= {13858\over 3 z_{12}^4 z_{13}^4 z_{23}^4} }

\subsec{The $k=2$ extremal CFT}

The $k=2$ extremal CFT, if exists, has 1-loop partition function
\eqn\partkt{Z_2(q) = J(q)^2 - 393767, } By comparing with the
six-point function of the twist operator ${\cal E}$, in particular, the three-point function
$\langle \Psi_x(0) \Psi_y(u) \Psi_z(v) \rangle$ up to order ${\cal O}(x^{-2} y^{-2} z^0)$, we can uniquely
fix the genus two partition function, \eqn\gteoket{\eqalign{ &Z_{k=2,g=2}(\Omega) =
{A_2\over \chi_{10}^2}\left( {574489\over 12230590464}\psi_4^6 +
{1125863\over 1528823808}\psi_4^3 \psi_6^2 + {159769\over 764411904}
\psi_6^4 \right.\cr & \left.-{17809159\over 905969664} \psi_4^3
\chi_{12} -{6550529\over 226492416}\psi_6^2\chi_{12}  +
{91785533041\over 154618822656} \chi_{12}^2\right. \cr & \left.
-{393767\over 1572864} \psi_4^2 \psi_6 \chi_{10} +
{229938936071\over 9663676416} \psi_4 \chi_{10}^2 \right) }} This
partition function has the correct singular behavior as
$e_{12},e_{34},e_{56}\to 0$. Furthermore, as $\epsilon \to 0$, \gteoket\ indeed factorizes as \eqn\facte{ Z_{k=2,g=2}(\Omega)
\to {const\over \epsilon^4} Z_2(\tau_1) Z_2(\tau_2). } This is a
highly nontrivial consistency check of \gteoket, which was determined without
implementing \facte.

Similarly to the $k=1$ case, we can expand the six point function \sixpo\ corresponding to \gteoket,
up to order ${\cal O}(x^0 y^0 z^0)$, and extract information about the three-point function of the primaries of dimension 3. There are
42987519 such primaries, denoted by ${\cal O}_i$, whose two-point functions
are normalized as before. We find
\eqn\theoa{ {1\over 42987519} \sum_{i,j,k=1}^{42987519} \langle{\cal O}_i(z_1) {\cal O}_j(z_2) {\cal O}_k(z_3)\rangle^2
= {104725\over 4 z_{12}^6 z_{13}^6 z_{23}^6} }
As a piece of numerology, the fact that \theoa\ is almost an integer multiple of $z_{12}^{-6} z_{13}^{-6}z_{23}^{-6}$ suggests that all the 
dimension 3 primaries ${\cal O}_i$ may be in one irreducible representation of some symmetry group (possibly containing the monster group as a subgroup).

\subsec{The $k=3$ extremal CFT}

The $k=3$ extremal CFT has 1-loop partition function \eqn\partfn{
Z_3(q) = J(q)^3-590651 J(q) - 64481279. } The genus two partition function is $1/\chi_{10}^3$ times
a weight 36 Siegel modular form. There are 17 independent Siegel modular forms of weight 36: $\psi_4^9$, $\psi_4^6\psi_6^2$, $\cdots$,
$\chi_{10}^3\psi_6$. It turns out that by comparing with the six-point function of the twist operator ${\cal E}$ \sixpo,
up to the terms of order ${\cal O}(x^{-3}y^{-3}z^{-1})$, which does not require the knowledge of correlation functions of the
dimension 4 primaries, we can determine all but 3 linear combinations of the 17 coefficients. The remaining 3 coefficients can be
fixed by demanding factorization in the limit
 $\epsilon\to 0$,
\eqn\aafs{ Z_{k=3,g=2} \to {const\over \epsilon^6} Z_3(\tau_1) Z_3(\tau_2). }
This is not obviously possible, since the factorization a priori over-determines the remaining 3 coefficients.
Remarkably, we do find a unique and consistent solution:
\eqn\udntenewfinal{ \eqalign{ & Z_{k=3,g=2}(\Omega) = {A_3\over \chi_{10}^3}\left[ {307082041\over 1352605460594688}\psi_4^9
+ {1025849351\over 112717121716224} \psi_4^6\psi_6^2 + {579427513\over 28179280429056} \psi_4^3 \psi_6^4\right. \cr & \left.
+ {36867719\over 21134460321792} \psi_6^6 - {9519543271\over 66795331387392} \psi_4^6 \chi_{12}
-{15531189821\over 8349416423424} \psi_4^3 \psi_6^2\chi_{12}\right. \cr & \left. - {1511576479\over 4174708211712} \psi_6^4\chi_{12}
+{328564579342237\over 17099604835172352} \psi_4^3 \chi_{12}^2
+{85316215289123\over 4274901208793088} \psi_6^2 \chi_{12}^2 \right. \cr & \left.
-{11321414397534479\over 60798594969501696} \chi_{12}^3
-{150649445\over 38654705664} \psi_4^5 \psi_6 \chi_{10}- {76160539\over 9663676416} \psi_4^2 \psi_6^3 \chi_{10}\right. \cr & \left.
+ {878731318367\over 1855425871872} \psi_4^2\psi_6 \chi_{10}\chi_{12}+{492299265760247\over 1068725302198272} \psi_4^4\chi_{10}^2 +
{256516494599113\over 267181325549568} \psi_4 \psi_6^2 \chi_{10}^2\right. \cr & \left.
-{36705982837911919\over 1266637395197952} \psi_4 \chi_{10}^2 \chi_{12} - {4272745361794189\over 118747255799808} \psi_6 \chi_{10}^3
\right]. } }
This can be regarded as a piece of evidence for the existence (and perhaps uniqueness) of the $k=3$ ECFT.
It would be also straightforward to extract the sum of squares of the three-point functions of dimension 4 primaries in the $k=3$ ECFT,
as in the $k=1,2$ cases; although,
we did not attempt this since the computation is rather time-consuming (even with our Mathematica program!).

\newsec{Factorization and sewing}

\subsec{ Next to leading order at the separating degeneration }

It is also possible to compute the less singular terms in the
expansion in $\epsilon$ near the separating degeneration. In
practice, the expansion is easier to set up by working with the six
point functions of twist fields. In the limit where three twist
fields ${\cal E}$  are brought together and replaced by a generic
operator in the  twisted sector, the six point function factorizes
into two four point functions, each corresponding to a torus
partition function. A four point function with a generic operator in
the twisted sector $O^1_{-n_1/2}O^2_{-n_2/2}\cdots |{\cal E}\rangle$
roughly corresponds to a torus one point function of
$O^1_{-n_1}O^2_{-n_2}\cdots |0\rangle$.  For an extremal CFT the
second nonzero operator in the twisted sector after ${\cal E}$ is
$L_{-1}\cdot{\cal E}(z)=-\partial {\cal E}(z)$.

The factorization limit can be set up, for example, as the $t\to 0$
limit of \eqn\fafsix{\langle {\cal E}(t e_1) {\cal E}(t e_2) {\cal
E}(t e_3){\cal E}(1/f_1) {\cal E}(1/f_2) {\cal E}(1/f_3) \rangle.}
For convenience we will choose $e_1+e_2+e_3=f_1+f_2+f_3=0$.
The leading term in the six-point function, of order ${\cal O}(t^{-3k})$,
will be \eqn\tsz{\eqalign{ & t^{-3k} \langle {\cal
E}(e_1) {\cal E}(e_2) {\cal E}(e_3){\cal E}'(\infty) \rangle \langle
{\cal E}(0){\cal E}(1/f_1) {\cal E}(1/f_2) {\cal E}(1/f_3) \rangle
\cr &= t^{-3k} {Z_{g=1}(\tau_1)Z_{g=1}(\tau_2)\over (e_{12}e_{23}e_{13})^k (\tilde f_{12}\tilde f_{23}\tilde f_{13})^k} }}
where ${\cal E}'$ stands for the operator ${\cal E}$ in the $u=1/z$ frame,
$\tau_1=\tau(e_1,e_2,e_3,\infty)$, $\tau_2=\tau(f_1,f_2,f_3,\infty)$, as in \jgen, and $\tilde f_{ij}
\equiv f_i^{-1}-f_j^{-1}$.
The first subleading term in \fafsix, of order ${\cal O}(t^{1-3k})$, will be \eqn\subt{\eqalign{&{t^{1-3k} \over 3k } \langle
{\cal E}(e_1) {\cal E}(e_2) {\cal E}(e_3)(L_{-1}\cdot{\cal E})'(\infty)
\rangle \langle L_{-1}\cdot{\cal E}(0){\cal E}(1/f_1) {\cal E}(1/f_2)
{\cal E}(1/f_3) \rangle \cr &= {t^{1-3k}\over 3k}\left[ {\left.\partial_x Z_{g=1}(\tau(e_i,1/x))\right|_{x=0}
\over (e_{12}e_{23}e_{13})^k}\right] \left[{(f_1f_2f_3)^{3k}\left.\partial_x Z_{g=1}(\tau(f_i,1/x))\right|_{x=0}\over (f_{12}f_{23}f_{13})^k}
\right]
\cr &={t^{1-3k}\over 3k} {{1\over 2\pi i}\partial_\tau Z_{g=1}(\tau_1)\,{1\over 2\pi i}\partial_\tau Z_{g=1}(\tau_2)\over
(e_{12}e_{23}e_{13})^k (\tilde f_{12}\tilde f_{23}\tilde f_{13})^k}, }} where
the factor $1/3k$ comes from the normalization of $L_{-1}|{\cal E}\rangle$, and we have used
the identity
 \eqn\jstau{\eqalign{
\left.\partial_x\tau(e_i,1/x)\right|_{x=0} &= {2^{8}\cdot27
e_1e_2e_3(e_1^2+e_2^2+e_3^2-e_1e_2-e_2e_3 -e_3e_1)^2\over
(e_{12}e_{23}e_{13})^2 \, \partial_\tau j(\tau_1)} \cr
&=-{E_6(\tau_1)\over E_4(\tau_1)} {j(\tau_1)\over\partial_\tau
j(\tau_1)}={1\over 2\pi i}. }} Note that ${1\over 2\pi
i}\partial_\tau Z(\tau)$ is the torus one-point function of the
stress-energy tensor. In the examples of $k=1,2,3$, one can rewrite
the genus two partition functions \thea, \gteoket, \udntenewfinal\
in the form of the six-point function \fafsix\ using \basiss,
\siegelgen, and expand in $t$. The result indeed matches \tsz,
\subt\ precisely. Note that $t$ is related to the parameter
$\epsilon$ of \refs{\Tuite,\Tuiteb} by $t\sim\epsilon^4$.

\subsec{ Genus two partition function from sewing tori }

Generally, the genus two partition function of a holomorphic CFT of
central charge $c=24k$ with small $\epsilon$ (as defined in
\refs{\Tuite,\Tuiteb}) can be expanded as \eqn\patge{
Z_{g=2}(\tau_1,\tau_2,\epsilon) = \sum_i \epsilon^{2\Delta_i-2k}
\langle {\cal A}_i\rangle_{\tau_1} \langle {\cal
A}_i\rangle_{\tau_2} } where ${\cal A}_i$ are an orthonormal basis
of operators, with dimension $\Delta_i$, $\langle
\cdots\rangle_\tau$ stands for the one-point function on a torus of
modulus $\tau$, with $\langle 1\rangle_\tau=Z_{g=1}(\tau)$. For an
ECFT, all the operators with $\Delta\leq k$ are Virasoro descendants
of 1, and their torus one-point functions can be derived using Ward
identities. One can also constrain the torus one-point function of a
general primary field ${\cal O}$ of dimension $\Delta(>0)$. It can
be written as \eqn\onepot{ \langle {\cal O}\rangle_\tau = {\rm Tr}
{\cal O}_0 q^{L_0-k}, } where ${\cal O}_0 =\oint {dz\over 2\pi i}
{\cal O}(z)$ is the zero mode of ${\cal O}$. Furthermore, $\langle
{\cal O}\rangle_\tau$ is a modular form of weight $\Delta$. The
trace in \onepot\ does not receive contribution from Virasoro
descendants of the vacuum, and hence the leading term in the
$q$-expansion of \onepot\ is of order $q^{(k+1)-k}=q$. Therefore
$\langle{\cal O}\rangle_\tau$ is a cusp form of weight $\Delta$, and
can be non-vanishing only for $\Delta\geq 12$. We will not attempt
to further constrain $\langle {\cal O}\rangle_\tau$, which requires
the knowledge of three-point functions of the primaries.

If $k\geq 11$, we can in principle determine the singular part as
well as the ${\cal O}(1)$ part of $Z_{g=2}(\tau_1,\tau_2,\epsilon)$
in the $\epsilon\to 0$ limit. If $k\leq 10$, we know that the torus
one-point functions of the primaries with $\Delta\leq 11$ vanish,
hence knowing the one-point function of the Virasoro descendants of
1, up to dimension 11, we can in principle fix the terms in
$Z_{g=2}(\tau_1,\tau_2,\epsilon)$ up to ${\cal
O}(\epsilon^{2(11-k)})$.

On the other hand, by modular invariance we expect $Z_{g=2}$ to take
the general form \eqn\tsmod{ Z_{g=2}(\Omega) = \sum_{m=0}^{\lfloor
6k/5\rfloor} \chi_{10}^{-k+m} P_{12k-10m}(\psi_4,\psi_6,\chi_{12}) }
where $P_{12k-10m}(\psi_4,\psi_6,\chi_{12})$ is a polynomial in
$\psi_4,\psi_6,\chi_{12}$ of homogeneous weight $12k-10m$. The
leading terms in the expansion of
$\psi_4,\psi_6,\chi_{10},\chi_{12}$ in the small $\epsilon$ limit
are given in \degensie. If we know the terms of order ${\cal
O}(\epsilon^{-2k+2m})$ in \patge, the polynomials $P_{12k-10m}$ are
fixed correspondingly.

For example, the leading singularity $\epsilon^{-2k}$ comes from the
term $\chi_{10}^{-k} P_{12k}$, \eqn\otse{
\epsilon^{-2k}(\Delta(\tau_1)\Delta(\tau_2))^{-k}
P_{12k}\left({1\over 4}E_4(\tau_1) E_4(\tau_2),{1\over
16}E_6(\tau_1) E_6(\tau_2), 96 \Delta(\tau_1)\Delta(\tau_2)\right) }
Writing \eqn\xydef{ {E_4(\tau_i)^3\over
1728\Delta(\tau_i)}=x_i,~~~~{E_6(\tau_i)^2\over
1728\Delta(\tau_i)}=x_i-1,~~~~i=1,2, } \otse\ can be put in the form
\eqn\sint{ \epsilon^{-2k} \tilde P(x_1 x_2,(x_1-1)(x_2-1)) } for
some polynomial $\tilde P$. On the other hand, the leading term in
\patge\ is of the form \eqn\trs{ \epsilon^{-2k} H(x_1) H(x_2), } for
some polynomial $H(x)$, since $Z_{g=1}(\tau_i)$ is a polynomial in
$J(\tau_i)=1728x_i-744$. There is a unique way of rewriting \trs\ in
the form \sint, which determines $P_{12k}(\psi_4,\psi_6,\chi_{12})$.

Similarly, comparison with the subleading terms in $\epsilon$ in
\patge\ will in principle\foot{A complication lies in the expansion
of the Siegel modular forms in $\epsilon$, which may be obtained
using the formulae in \refs{\Tuiteb}. } determine $P_{12k-10}$,
$P_{12k-20}$, $\cdots$. For $k\geq 11$, this will fix the $P$'s up
to $P_{2k}$. The remaining $P_{2k-10}$, $\cdots$,
$P_{2k-10\lfloor{k\over 5}\rfloor}$ are not determined in this
approach. For $k\leq 10$, since one can determine the terms in
\patge\ up to ${\cal O}(\epsilon^{2(11-k)})$, all the polynomials
$P_{12k-10m}$ are fixed. Therefore the genus two partition functions
of the ECFTs with $k\leq 10$ are in principle uniquely fixed. To
check the consistency of these partition functions (which is not a
priori obvious), one should consider the limit where a handle
pinches, say by comparing with the six-point function of twist
fields as discussed in previous sections, or with two-point
functions on the torus (the self-sewing of \Tuiteb). The consistency
checks and explicit computations are left to future work.

\bigskip

\centerline{\bf Acknowledgement} We are grateful to D. Shih and A.
Strominger for discussions, and especially to E. Witten for
correspondences and detailed comments on an earlier draft of the
paper. DG is supported in part by DOE grant DE-FG02-91ER40654. XY is
supported by a Junior Fellowship from the Harvard Society of
Fellows.

\bigskip

\listrefs

\end